\documentclass[aps,twocolumn]{revtex4}
\usepackage{epsfig}
\usepackage{graphicx}
\usepackage{amsmath,amssymb,amsfonts}
\usepackage{array}
\usepackage{url}
\usepackage{hyperref}
\usepackage{multirow}
\usepackage{float}
\usepackage{lineno}
\usepackage{xspace}
\usepackage{ulem}
\usepackage[usenames,dvipsnames]{color}

\newcommand{\bef}{\begin{figure}}
\newcommand{\eef}{\end{figure}}
\newcommand{\bc}{\begin{center}}
\newcommand{\ec}{\end{center}}

\newcommand{\be}{\begin{equation}}
\newcommand{\ee}{\end{equation}}
\newcommand{\bea}{\begin{eqnarray}}
\newcommand{\eea}{\end{eqnarray}}

\def\ba{\begin{eqnarray}}
\def\ea{\end{eqnarray}}
\begin{document}
\title{Proton number cumulants in a modified van der Waals hadron resonance gas}
 
\author{Kshitish Kumar Pradhan}
\author{Ronald Scaria}
\author{Dushmanta Sahu}
\author{Raghunath Sahoo\footnote{Corresponding Author: Raghunath.Sahoo@cern.ch}}
\affiliation{Department of Physics, Indian Institute of Technology Indore, Simrol, Indore 453552, India}

\begin{abstract}

An estimate of the proton number cumulants in the hadronic matter is presented considering a van der Waals-type interaction between the constituent particles. We argue that the attractive and repulsive parameters in the VDW hadron resonance gas (VDWHRG) model change as functions of baryochemical potential ($\mu_{B}$) and temperature ($T$). This, in turn, affects the estimation of thermodynamic properties and, consequently, the conserved charge fluctuations. We employ a simple parametrization to bring in the center-of-mass energy ($\sqrt{s_{\rm NN}}$) dependence on temperature and baryochemical potential and then estimate the proton number cumulants with the modified approach. The modified van der Waals hadron resonance gas model (MVDWHRG) explains the existing experimental data very well.

\end{abstract}
\date{\today}
\maketitle

\section{Introduction}
\label{intro}

Exploring the quantum chromodynamics (QCD) phase diagram is of interest to the experimental and theoretical high-energy nuclear physics community. The existence of two phases, the hadronic phase and the quark-gluon plasma (QGP) phase, in this diagram, separated by a crossover transition at high temperatures ($T$) and low baryochemical potential ($\mu_B$), is now well accepted based on the lattice QCD (lQCD) calculations and other phenomenological models \cite{Bazavov:2017dus}. On the other hand, it is expected that a first-order phase transition exist between the hadronic and partonic matter at low $T$ and high $\mu_{B}$ regime. These phase transition lines meet at the hypothesized critical end-point (CEP). However, the applicability of lQCD breaks down at high $\mu_{B}$ due to the fermion sign problem \cite{Borsanyi:2013bia, HotQCD:2014kol}. Thus, to probe the CEP, which is a second-order phase transition \cite{Stephanov:1998dy, Stephanov:1999zu}, one needs to take the help of various phenomenological models and experiments.

It has been theorized \cite{Stephanov:1998dy, Stephanov:1999zu, Stephanov:2008qz, Hatta:2003wn}  that the higher-order cumulants of conserved charges, especially baryon number, are sensitive to the second-order phase transition. This is because the higher-order cumulants scale with the corresponding powers
of the correlation length; thus, a limited yet finite increase of the
correlation length due to the critical slowing down can still
be observed. Moreover, as the baryon number is a conserved
quantity, its fluctuations are mostly unmodified by the final-state
interactions in the hadronic phase. Some theoretical predictions demonstrate a non-monotonic behavior of the kurtosis of the baryon number distributions as a function of the collision energy, given that the chemical freeze-out happens close to the CEP \cite{Stephanov:2011pb}. In an experiment, the proton number is taken as a proxy for the baryon number given that the proton is the lightest and hence the most abundant baryon in the hadronic matter formed after a collision. An exploration towards higher $\mu_{B}$ by varying the collision energies in experiments to determine the possible existence of the CEP on the phase diagram is being carried out by the Beam Energy Scan II (BES-II) program at the Relativistic Heavy Ion Collider (RHIC).

On the phenomenological domain, an alternative to the lQCD approach at low $T$ (up to 150 MeV) is the ideal Hadron Resonance Gas (IHRG) model. The IHRG model has been observed to agree with the lQCD thermodynamic results for temperatures in this range at zero $\mu_{B}$ \cite{Bellwied:2013cta, HotQCD:2012fhj, Bellwied:2017ttj, Borsanyi:2011sw}. It can also successfully describe the particle ratios in high-energy collisions. Unlike lQCD, the IHRG model works very well even at very high $\mu_{\rm B}$ regime. The IHRG model can thus be an excellent alternative to studying the low $T$ and high $\mu_{B}$ regime. 

The limiting condition of $T \leq$ 150 MeV for IHRG may be ignored while studying the thermodynamic properties. However, this must be addressed while estimating higher order charge fluctuations where the deviations from the lQCD estimates are substantial \cite{Bazavov:2013dta, Bazavov:2017dus, Borsanyi:2018grb, Schmidt:2012ka}. Recently, much focus has been diverted towards an interacting hadron resonance gas model as they extend the region of agreement with lQCD data. The excluded volume hadron resonance gas (EVHRG) model assumes an eigen volume parameter for the hadrons, which mimics a repulsive interaction between the hadrons \cite{Andronic:2012ut, Vovchenko:2014pka}. Meanwhile, the mean-field hadron resonance gas (MFHRG) model introduces a repulsive interaction through an interaction potential in the hadronic medium \cite{Kapusta:1982qd, Olive:1980dy}. There are also various other improvements to the HRG model in literature, such as the Lorentz modified excluded volume hadron resonance gas (LMEVHRG) model \cite{Pal:2020ink}, where the hadrons are treated as Lorentz contracted, and the effective thermal mass hadron resonance gas (THRG) model \cite{Zhang:2019uct}, where the hadrons gain effective mass with temperature. However, the most successful improvement which explains the lQCD results is the van der Waals hadron resonance gas (VDWHRG) model. This model assumes a van der Waals-type interaction among the hadrons, with long-range attractive and short-range repulsive interactions between the hadrons \cite{Vovchenko:2015xja, Vovchenko:2015pya, Vovchenko:2015vxa}. The VDWHRG effectively explains the lQCD data up to $T \simeq 180$ MeV \cite{Vovchenko:2016rkn, Vovchenko:2017cbu}. It may thus be inferred that van der Waals interaction does play a crucial role in hadronic systems at high temperatures. In addition, VDWHRG contains the possibility of a first-order liquid-gas phase transition, which is a part of the QCD phase diagram. This model has been used to estimate various thermodynamic and transport properties of the hadronic matter \cite{Samanta:2017yhh, Sarkar:2018mbk, Pradhan:2022gbm}. In addition, the conserved charge fluctuations have also been estimated in the ambit of the VDWHRG model \cite{Behera:2022nfn, Vovchenko:2016rkn}. However, very little focus has been given to the high $\mu_{B}$ regime.

In ref.~\cite{Vovchenko:2016rkn}, the authors show that the VDWHRG model explains the lQCD results better than HRG and EVHRG models. This is because repulsive interactions significantly suppress thermodynamic functions in the crossover region at zero $\mu_{B}$. In contrast, the attractive interactions result in an enhancement of the thermodynamic functions. The combination of both these interactions leads to better agreement of the model with lQCD results as compared to the other models. In ref.~\cite{Zhang:2019uct}, the authors update the VDWHRG model by considering the effective thermal mass of the hadrons. The liquid-gas critical point has also been explored by taking the VDWHRG approach in ref.~\cite{Samanta:2017yhh}. However, in all these studies, the authors have assumed the VDW interactions to be present only among (anti)baryons-(anti)baryons, and the meson-meson, meson-(anti)baryon interactions have been excluded for the sake of simplicity. In ref.~\cite{Sarkar:2018mbk}, a study on the criticality behavior has been conducted including both (anti)baryons-(anti)baryons and meson-meson interactions. The caveat is that in all of these studies \cite{Samanta:2017yhh, Sarkar:2018mbk}, the attraction and repulsion parameters, $a$ and $b$, are taken as constants for all $\mu_{B}$ and $T$, which seems to be an oversimplification.

In ref.~\cite{Vovchenko:2016rkn}, the authors have fixed the $a$ and $b$ parameters by reproducing the saturation density $n_{\rm 0}$ = 0.16 fm$^{-3}$ and binding energy $E/A = 16$ MeV of the ground state of nuclear matter. They obtain $a = 329$ MeV and $b = 3.42$ fm$^{3}$ \cite{Vovchenko:2016rkn, Vovchenko:2015pya, Redlich:2016dpb} and predict a first-order liquid-gas phase transition in the nuclear matter with a critical point at $T_{\rm c} \simeq 19.7$ MeV and $\mu_{\rm B} \simeq 908$ MeV. However, an alternate estimate of $a$ and $b$ may be obtained by fitting the lQCD data to study the behavior of a system formed in ultra-relativistic high-energy collisions \cite{Samanta:2017yhh}. The current study argues that the above assumptions are oversimplifications, and the attractive and repulsive parameters may depend on chemical potential and temperature. Using a chi-square minimization technique, we fit the $\mu_{\rm B}/T$ dependent lQCD data to obtain the $a$ and $b$ parameters at different $\mu_{B}/T$.  We further obtain a functional form for $a$ and $b$ as functions of $\mu_{\rm B}/T$, which can be used for any reasonably extrapolated $\mu_B$ and $T$ values. This essentially modifies the results of the VDWHRG model at a high baryon-density region. Taking this modified VDWHRG (MVDWHRG) approach, we try to explain the proton number fluctuation data from the STAR collaboration. 

The current work is organized as follows. In section \ref{formulation}, we briefly describe the van der Waals hadron resonance gas model. We then discuss the modification of the $a$ and $b$ parameters as functions of baryochemical potential. In section \ref{res}, we briefly discuss the obtained results and consequences. Finally, we summarise our results in section \ref{sum}.

\section{Formulation}
\label{formulation}

The ideal HRG formalism considers hadrons to be point particles with no interactions between them. Under this formalism, the partition function of $i^{th}$ particle species having mass $m_{i}$ is given in a grand canonical ensemble (GCE) as ~\cite{Andronic:2012ut},
\begin{equation}
\label{eq1}
ln Z^{id}_i = \pm \frac{Vg_i}{2\pi^2} \int_{0}^{\infty} p^2 dp\ ln\{1\pm \exp[-(E_i-\mu_i)/T]\},
\end{equation}
where the notations of $g_i$, $E_i = \sqrt{p^2 + m_i^2}$ and $\mu_i$ are the degeneracy, energy and chemical potential of the $i^{th}$ hadron, respectively. $\mu_i$ is further expanded in terms of the baryonic, strangeness, and charge chemical potentials ($\mu_{B}$, $\mu_{S}$ and $\mu_{Q}$, respectively) and the corresponding conserved numbers ($B_i$, $S_i$ and $Q_i$) as, 
\begin{equation}
\label{eq2}
\mu_i = B_i\mu_B + S_i\mu_S +Q_i\mu_Q,
\end{equation}

Pressure $P_i$, energy density $\varepsilon_i$, number density $n_i$, and entropy density $s_i$ in the ideal HRG formalism can now be obtained as, 
\begin{equation}
\label{eq3}
P^{id}_i(T,\mu_i) = \pm \frac{Tg_i}{2\pi^2} \int_{0}^{\infty} p^2 dp\ ln\{1\pm \exp[-(E_i-\mu_i)/T]\}
\end{equation}
\begin{equation}
\label{eq4}
\varepsilon^{id}_i(T,\mu_i) = \frac{g_i}{2\pi^2} \int_{0}^{\infty} \frac{E_i\ p^2 dp}{\exp[(E_i-\mu_i)/T]\pm1}
\end{equation}
\begin{equation}
\label{eq5}
n^{id}_i(T,\mu_i) = \frac{g_i}{2\pi^2} \int_{0}^{\infty} \frac{p^2 dp}{\exp[(E_i-\mu_i)/T]\pm1}
\end{equation}
\begin{align}
 s^{id}_i(T,\mu_i)=&\pm\frac{g_i}{2\pi^2} \int_{0}^{\infty} p^2 dp \Big[\ln\{1\pm  \exp[-(E_i-\mu_i)/T]\}\nonumber\\ 
&\pm \frac{(E_i-\mu_i)/T}{\exp[(E_i-\mu_i)/T]\pm 1}\Big].
 \label{eq6}
 \end{align}

 \begin{figure*}[ht!]
\begin{center}
\includegraphics[scale = 0.44]{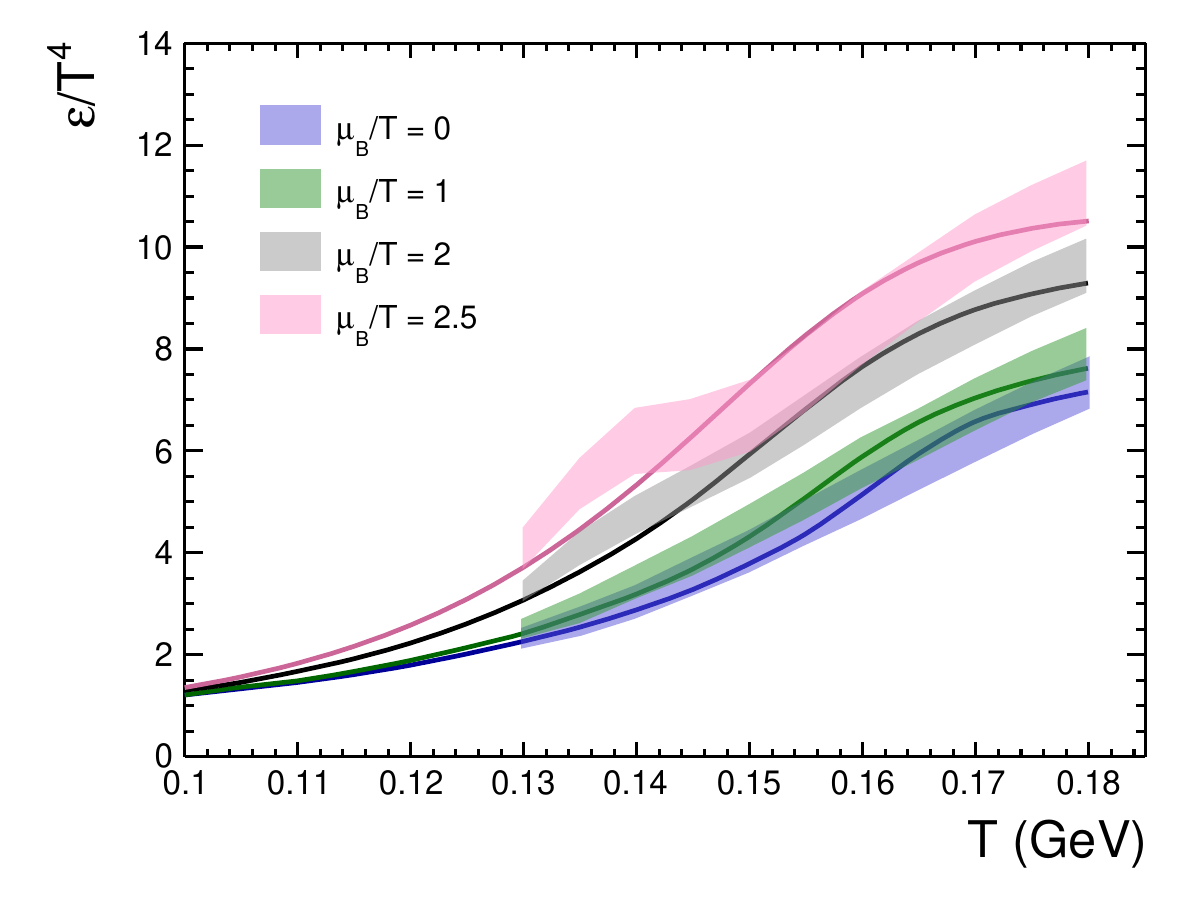}
\includegraphics[scale = 0.44]{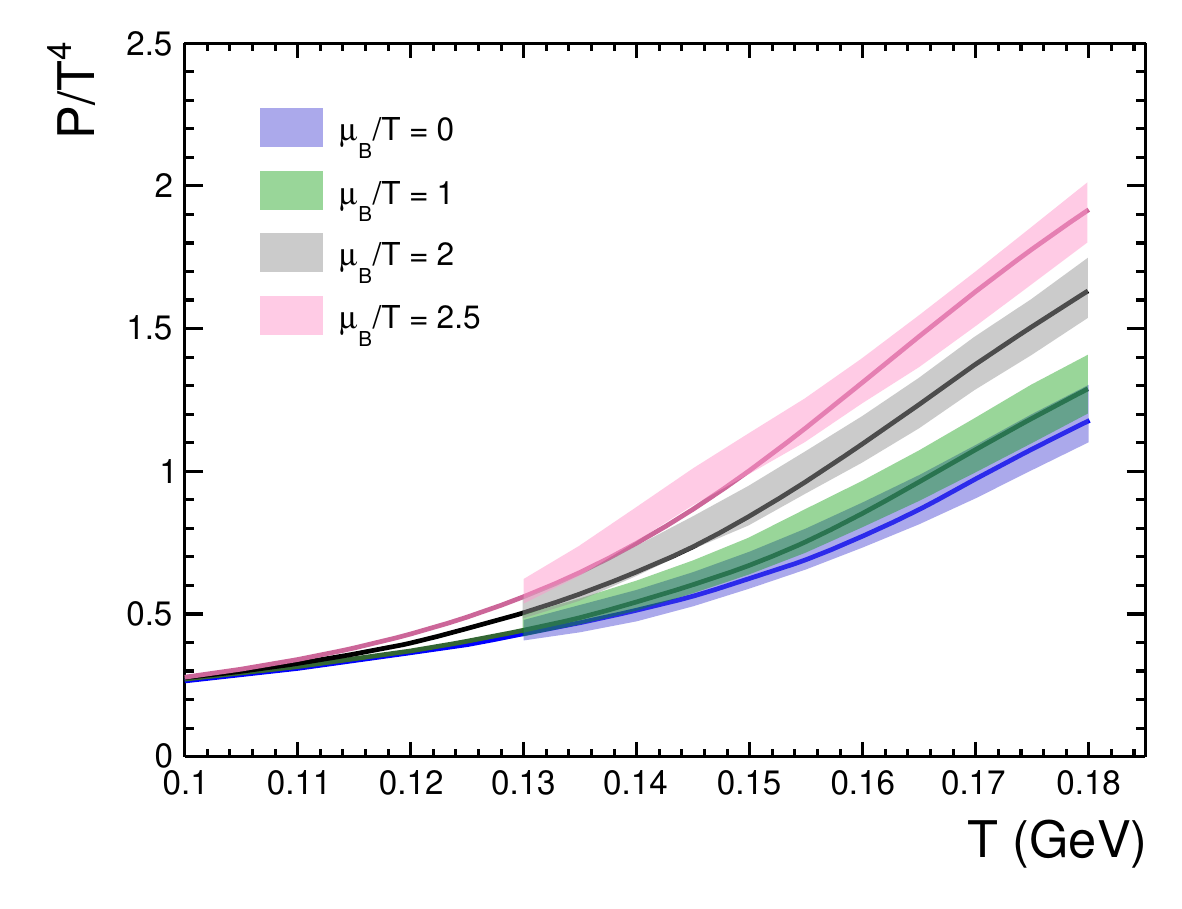}
8\caption{(Color Online) Simultaneous $\chi^{2}$ fit of scaled energy density and pressure at various $\mu_{B}/T$ as functions of temperature. The coloured bands are for lQCD data \cite{Bazavov:2017dus}, and the solid lines are the fits to the data.}
\label{fig1}
\end{center}
\end{figure*}

\begin{figure*}[ht!]
\begin{center}
\includegraphics[scale = 0.44]{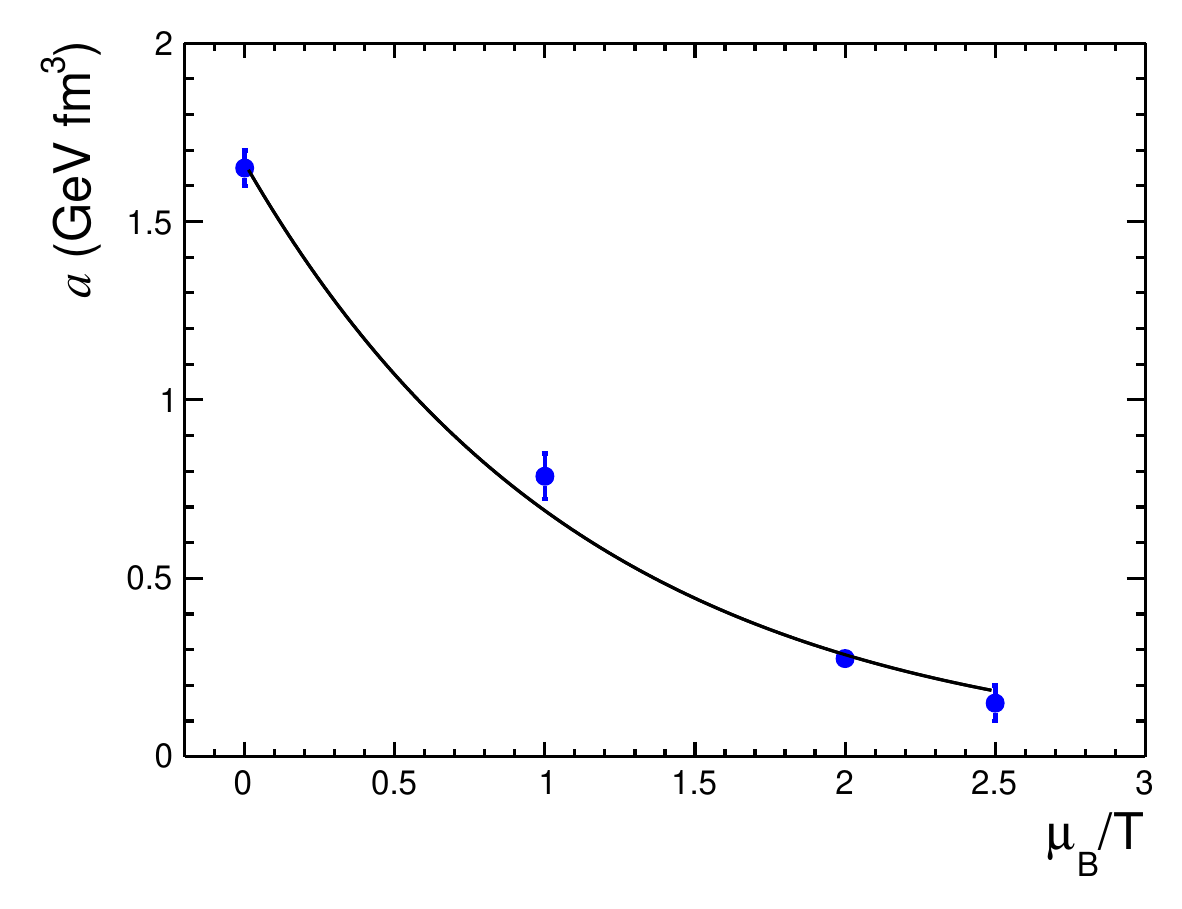}
\includegraphics[scale = 0.44]{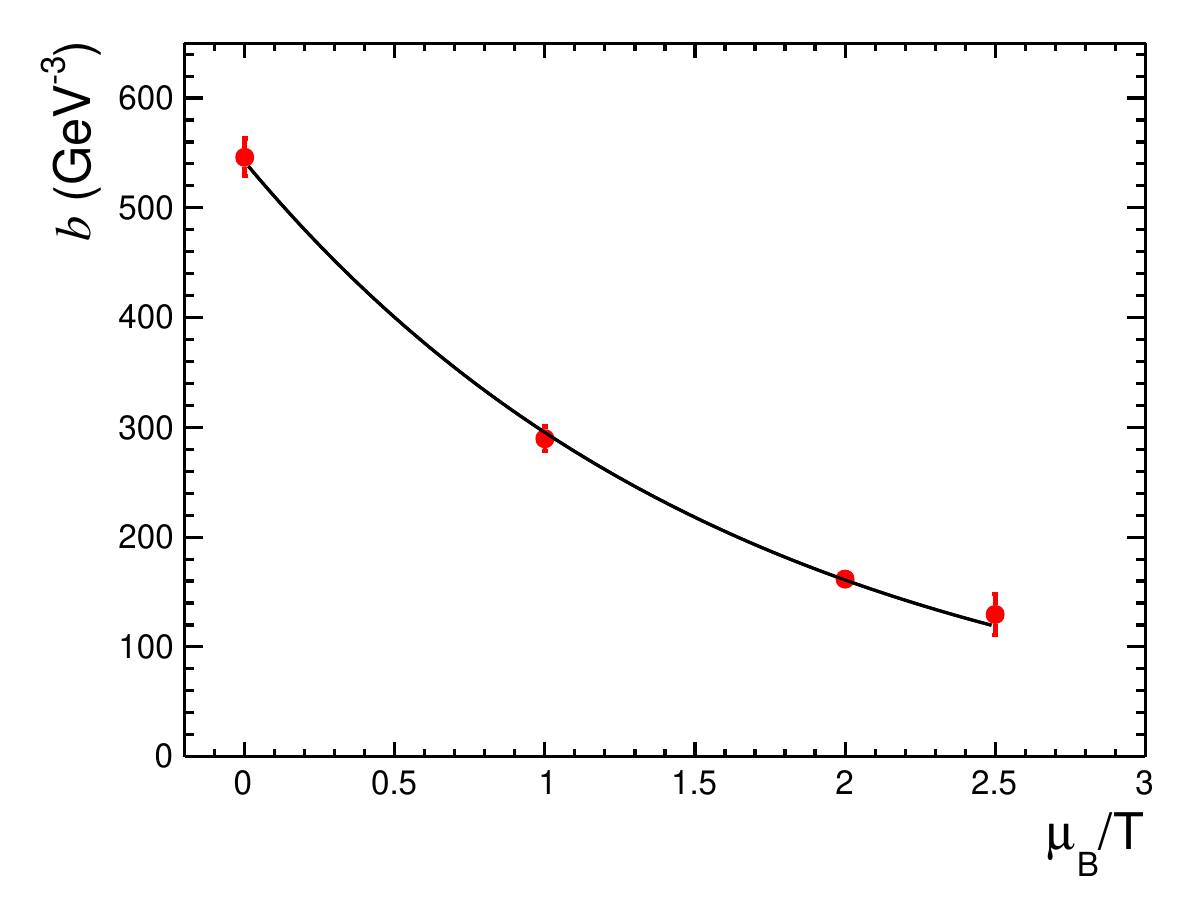}
\caption{(Color Online) Variation of extracted $a$ and $b$ parameters from the $\chi^{2}$ fit to the lQCD data as functions of $\mu_{B}/T$. The solid black lines denote the negative exponential fits to the $a$ and $b$ parameters.}
\label{fig2}
\end{center}
\end{figure*}

 The ideal HRG model can be modified to include van der Waals interactions between particles by the introduction of the attractive and repulsive parameters $a$ and $b$, respectively. This modifies the pressure and number density obtained in ideal HRG iteratively as follows \cite{Samanta:2017yhh, Vovchenko:2015vxa, Vovchenko:2015pya};
\begin{equation}
\label{eq9}
    P(T,\mu) = P^{id}(T,\mu^{*}) - an^{2}(T,\mu),
\end{equation}
where the $n(T,\mu$) is the VDW particle number density given by,
\begin{equation}
\label{eq10}
    n(T,\mu) = \frac{\sum_{i}n_{i}^{id}(T,\mu^{*})}{1+b\sum_{i}n_{i}^{id}(T,\mu^{*})}.
\end{equation}
Here, $i$ runs over all hadrons and $\mu^{*}$ is the modified chemical potential given by, 
\begin{equation}
\label{eq11}
    \mu^{*} = \mu - bP(T,\mu) - abn^{2}(T,\mu) + 2an(T,\mu).
\end{equation}
It is to be noted that the repulsive parameter is usually attributed to be related to the hardcore radius ($r$) of the particle through the relation $b = 16\pi r^{3}/3$, while no such definition is known for the attractive parameter.

Entropy density $s(T,\mu)$ and energy density $\epsilon(T,\mu)$ in VDWHRG can now be obtained as,
\begin{equation}
\label{eq12}
s(T,\mu) = \frac{s^{id}(T,\mu^{*})}{1+bn^{id}(T,\mu^{*})}
\end{equation}
\begin{equation}
\label{eq13}
\epsilon(T,\mu) = \frac{\sum_{i}\epsilon_{i}^{id}(T,\mu^{*})}{1+b\sum_{i}n_{i}^{id}(T,\mu^{*})} - an^{2}(T,\mu).
\end{equation}

Knowing the attractive and repulsive parameters, the total pressure in the VDWHRG model can be written as \cite{Samanta:2017yhh, Vovchenko:2015vxa, Vovchenko:2015pya, Vovchenko:2017cbu, Vovchenko:2016rkn, Sarkar:2018mbk}, 
\begin{equation}
\label{eq14}
P(T,\mu) = P_{M}(T,\mu) + P_{B}(T,\mu) + P_{\bar{B}}(T,\mu)
\end{equation}
where $M$, $B$ and $\bar B$ represents mesons, baryons and anti-baryons, respectively. Their contributions to pressure is now defined by,
\begin{equation}
\label{eq15}
P_{M}(T,\mu) = \sum_{i\in M}P_{i}^{id}(T,\mu^{*M})       
\end{equation}
\begin{equation}
\label{eq16}
P_{B}(T,\mu) = \sum_{i\in B}P_{i}^{id}(T,\mu^{*B})-an^{2}_{B}(T,\mu)
\end{equation}
\begin{equation}
\label{eq17}
P_{\bar{B}}(T,\mu) = \sum_{i\in \bar{B}}P_{i}^{id}(T,\mu^{*\bar{B}})-an^{2}_{\bar{B}}(T,\mu).
\end{equation}
The excluded volume correction modifies the meson chemical potential as $\mu^{*M}$, while VDW interactions give  modified chemical potentials of baryons and anti-baryons as $\mu^{*B}$ and $\mu^{*\bar B}$ ~\cite{Sarkar:2018mbk}. For the simple case of vanishing electric charge and strangeness chemical potentials \cite{Bazavov:2017dus}, $\mu_{Q} = \mu_{S} = 0$, these modified chemical potentials are obtained from Eq.~\ref{eq2} and Eq.~\ref{eq11} as; 
\begin{equation}
\label{eq18}
\mu^{*M} = -bP_{M}(T,\mu)
\end{equation}
\begin{equation}
\label{eq19}
\mu^{*B(\bar B)} = \mu_{B(\bar B)}-bP_{B(\bar B)}(T,\mu)-abn^{2}_{B(\bar B)}+2an_{B(\bar B)}
\end{equation}
where $n_{M}$, $n_{B}$ and $n_{\bar B}$ denotes modified number densities of each type of particle given by
\begin{equation}
\label{eq20}
    n_{M}(T,\mu) = \frac{\sum_{i\in M}n_{i}^{id}(T,\mu^{*M})}{1+b\sum_{i\in M}n_{i}^{id}(T,\mu^{*M})}
\end{equation}
\begin{equation}
\label{eq21}
    n_{B(\bar B)}(T,\mu) = \frac{\sum_{i\in B(\bar B)}n_{i}^{id}(T,\mu^{*B(\bar B)})}{1+b\sum_{i\in B(\bar B)}n_{i}^{id}(T,\mu^{*B(\bar B)})}.
\end{equation}
Here the summation runs over mesons ($M$), baryons ($B$), and anti-baryons ($\bar B$) respectively.

As already discussed, the VDW parameters $a$ and $b$ may depend upon temperature and chemical potential. 
In ref.~\cite{Dutra:2020qsn}, the authors have considered a density-dependent VDW model where the VDW parameters are no longer constants but are defined as functions of density. Taking forward the assumption of a fixed hard core radius for mesons, we have estimated the VDW parameters for (anti)baryon-(anti)baryon interactions for different $\mu_B/T$ values by fitting the corresponding lQCD energy density and pressure~\cite{Bazavov:2017dus}. In order to extract the van der Waals parameters $a$ and $b$, we use the $\chi^2$ minimization technique to obtain the best fit to the available lQCD
data ~\cite{Bazavov:2017dus} using the relation \cite{Samanta:2017yhh, Vovchenko:2014pka},
\begin{equation}
\label{chi2m}
\chi^2 = \frac{1}{N_{df}}\sum_{i,j}\frac{\big(X^{lQCD}_{i,j}(T_j)-X^{model}_{i,j}(T_j)\big)^2}{\big(\Delta^{lQCD}_{i,j}(T_j)\big)^2} ,
\end{equation}
where the $X^{lQCD}_{i,j}(T_j)$ and $X^{model}_{i,j}(T_j)$ are the values of the $i^{th}$ thermodynamic observable from lQCD calculations and model calculations respectively at a given $j^{th}$ temperature. $\Delta^{lQCD}_{i,j}$ is the error involved in the lQCD calculation. $N_{df}$ is the number of lQCD data points minus the number of fitting parameters. Fig. \ref{fig1} shows the goodness of the fit (solid line) to the lQCD data (color band) for different values of $\mu_{B}/T$ for scaled pressure and energy density. 
The obtained parameters and the corresponding $\chi^2$ values for each case of $\mu_{B}/T$ are listed in Table \ref{table1}.
The attractive and repulsive parameters, $a$ and $b$, are now shown as a function of $\mu_{B}/T$ in fig.~\ref{fig2}. It is observed that both these parameters decrease as functions of $\mu_{B}/T$. Therefore an increase in the baryon density or a decrease in the temperature leads to the weakening of the van der Waals interaction. This behavior is similar to what was observed in ref \cite{Dutra:2020qsn}. To parameterize these two parameters as functions of $\mu_{B}$ and $T$, the obtained $a$ and $b$ values are fitted with a negative exponential function. 
Therefore the van der Waals parameters now vary as functions of $\mu_{B}/T$, quantified by the relation,
\begin{equation}
    \begin{split}
    \label{par_expo}
    a = c_{1}\exp(c_{2}\frac{\mu_{B}}{T}),\\
    b = d_{1} \exp(d_{2}\frac{\mu_{B}}{T}).
    \end{split}
\end{equation}
where the constants $c_{1}$, $c_{2}$, $d_{1}$, and $d_{2}$ are given by, $c_{1}$ = 1.66 $\pm$ 0.05 GeV fm$^{3}$, $c_{2}$ = -0.88 $\pm$ 0.04, $d_{1}$ = 541.93 $\pm$ 15.98 GeV$^{-3}$, and $d_{2}$ = -0.61 $\pm$ 0.03. As it is evident from fig.~\ref{fig2}, both the VDW parameters are decreasing exponentially with $\mu_{B}/T$, therefore at high baryon density and low temperature the interacting hadron resonance gas approaches towards the ideal gas condition. In view of the van der Waals parameters getting modified with $\mu_B/T$, the model is henceforth referred to as the modified van der Waals HRG (MVDWHRG) model.

\begin{table}[h!]
\begin{center}
\begin{tabular}{ |c|c|c|c|c| } 
\hline
$\mu_{B}/T$ & $a$ (GeV fm$^3$) & $r_{B}$ (fm) & $\chi^2$\\[0.5ex]
\hline\hline
 0.0 & 1.650 $\pm$ 0.05 & 0.635 $\pm$ 0.05 & 1.06/20 \\ 
 1.0 & 0.786 $\pm$ 0.064 & 0.515 $\pm$ 0.05 & 0.91/20 \\ 
 2.0 & 0.275 $\pm$ 0.025 & 0.425 $\pm$ 0.05 & 1.88/20 \\ 
 2.5 & 0.150 $\pm$ 0.05 & 0.385 $\pm$ 0.15 & 3.5/20 \\ 
\hline
\end{tabular}
\caption{van der Waals parameters obtained using $\chi^2$ minimization technique for different $\mu_{B}/T$ lQCD data. Here $r_{\rm B}$ is the baryon radius.}
\label{table1}
\end{center}
\end{table}

\begin{figure*}[ht!]
\begin{center}
\includegraphics[scale = 0.29]{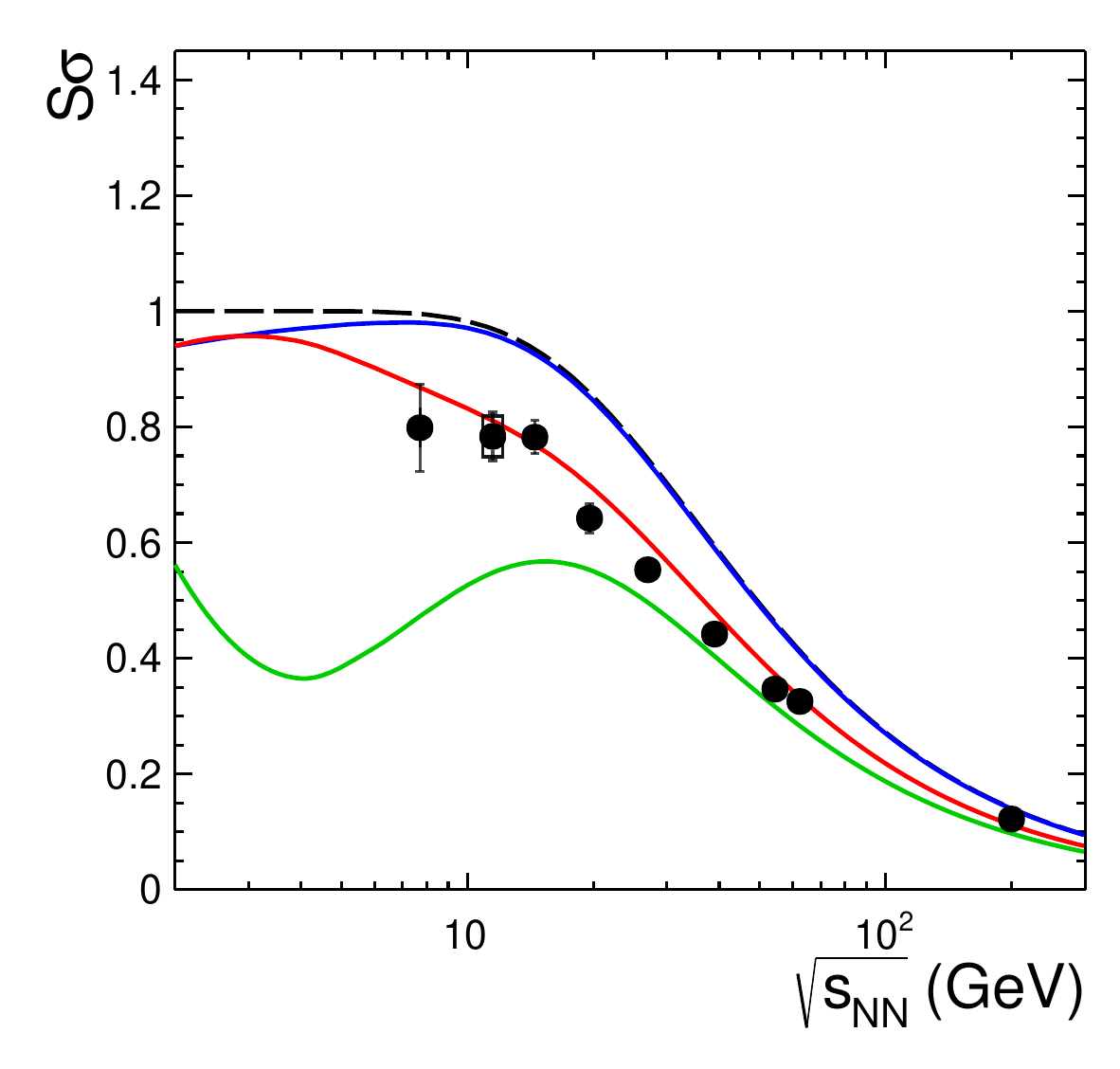}
\includegraphics[scale = 0.29]{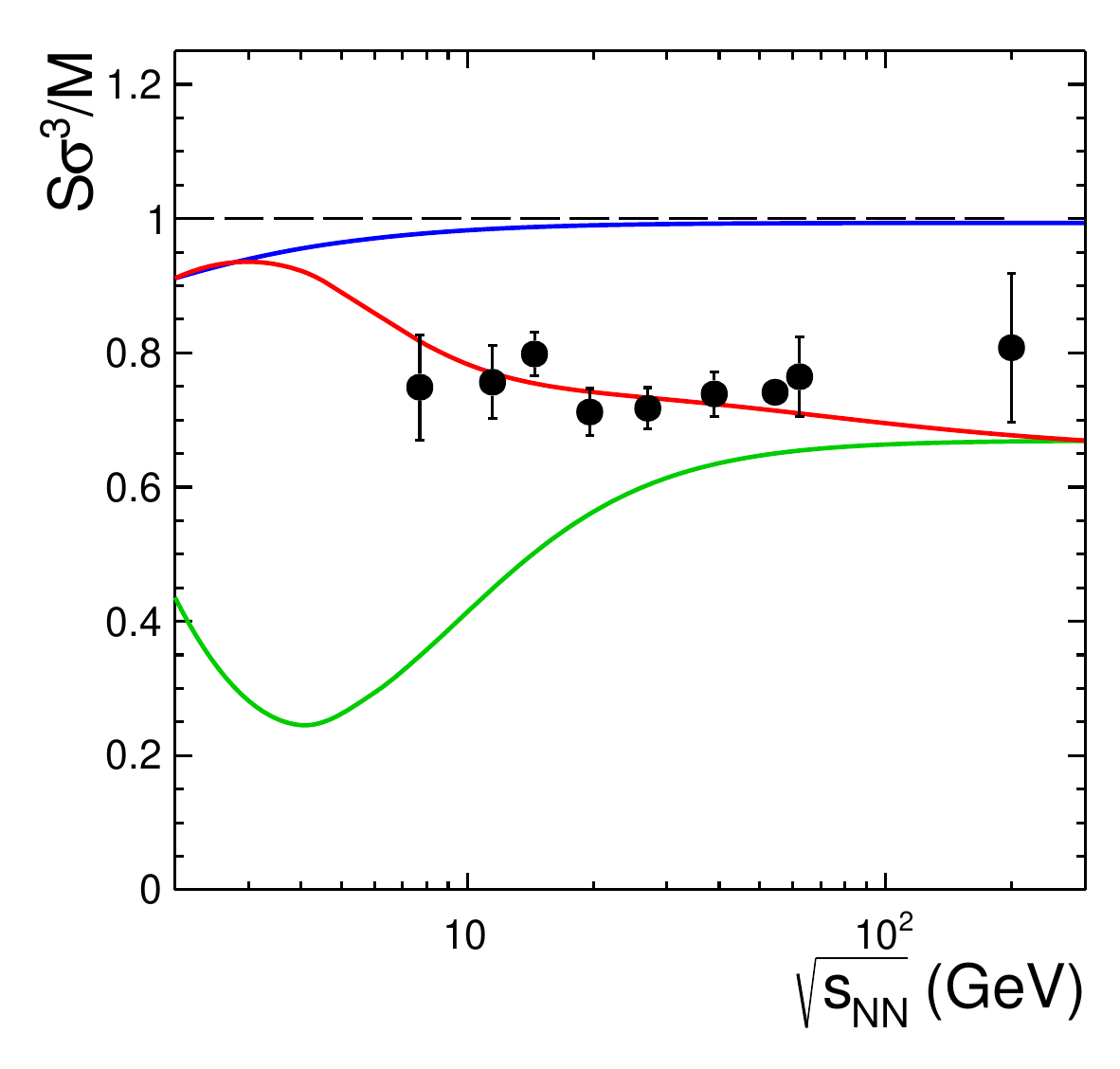}
\includegraphics[scale = 0.29]{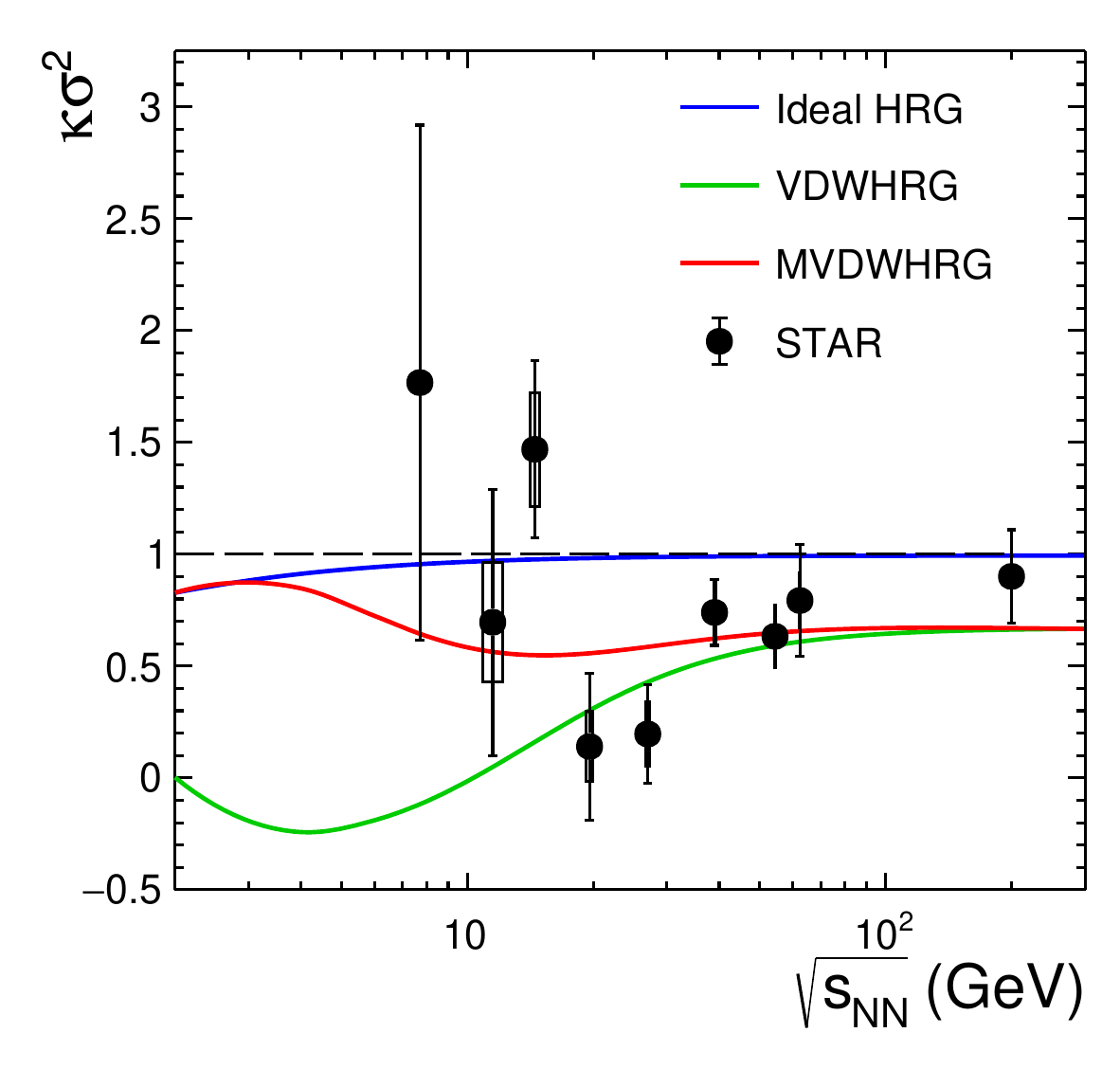}
\caption{(Color Online) Dimensionless ratios of net proton cumulants as functions of center-of-mass energies. The blue, green, and red solid lines represent results obtained with IHRG, VDWHRG and MVDWHRG respectively. The obtained results are compared with the STAR data \cite{STAR:2021iop} represented by black markers.}
\label{fig3}
\end{center}
\end{figure*}

The fluctuations of conserved charges can indicate the existence of a second-order phase transition and hence the position of the critical point in the QCD phase diagram. Such a point is a thermodynamic singularity where the susceptibilities diverge, and the order parameter fluctuates \cite{Stephanov:1999zu}. Therefore the susceptibilities and their ratios show non-monotonic behavior as a function of collision energy. Fluctuations of a conserved quantity and its higher moments can be obtained from the derivatives of the basic thermodynamic quantity, pressure, with respect to the corresponding chemical potential. These derivatives are known as susceptibilities or cumulants. The net proton number fluctuations can be studied as a proxy for the net baryon number fluctuations. The susceptibilities are given by,

\begin{equation}
    \label{suscep}
    \kappa_{n} = \frac{\partial^n}{\partial (\mu_{B}/T)^n}\frac{P}{T^4}.
\end{equation}
where the $P$ is the pressure defined in Eq.~\ref{eq3} and with the inclusion of van der Waals interaction, it is given by Eq.~\ref{eq14}. From Eq.~\ref{suscep}, the first derivative gives the mean of the net proton number as, $M = VT^{3}\kappa_{1}$, where $\kappa_{1}$ is simply the net proton number density given by $(n_{p}-n_{\bar p})/T^{3}$. Similarly, one can define the variance (Gaussian width), $\sigma^{2}$ along with the non-Gaussian fluctuations such as the skewness, $S$, and the kurtosis, $\kappa$ \cite{Stephanov:2008qz, Karsch:2010ck, Fukushima:2014lfa, Vovchenko:2021kxx}, so as to define the cumulant ratios as,
\begin{equation}
    \label{ratios}
    S\sigma = \frac{\kappa_3}{\kappa_2}, \hspace{0.5cm} \frac{S\sigma^3}{M} = \frac{\kappa_3}{\kappa_1}, \hspace{0.5cm}   \kappa\sigma^2 = \frac{\kappa_4}{\kappa_2}.
\end{equation}

We can now estimate these ratios using the MVDWHRG model. However, in order to compare our results with the experimentally available data, we need to relate the parameters of the hadron resonance gas, $\mu_{B}$ and $T$ with the collision energy, $\sqrt{s_{\rm NN}}$. Such a parametrization is available in literature~\cite{Cleymans:2005xv} and is given by,
\begin{equation}
    \label{paramet1}
    T(\mu_{B}) = q_{1} - q_{2}\mu_{B}^{2} - q_{3}\mu_{B}^{4},
\end{equation}
\begin{equation}
    \label{paramet2}
    \mu_{B}(\sqrt{s_{NN}}) = \frac{q_{4}}{1+q_{5}\sqrt{s_{NN}}},
\end{equation}

where the fitting parameters are given as $q_1$ = 0.166 GeV, $q_2$ = 0.139 GeV$^{-1}$, $q_{3}$ = 0.053 GeV$^{-3}$, $q_{4}$ = 1.308 GeV, and $q_{5}$ = 0.273 GeV$^{-1}$. Note that these parameters are obtained by using a freeze-out criterion based on the IHRG model and are not quite model dependent \cite{Cleymans:2005xv}. There is a very minimal change in the parameters when including interactions like excluded volume (EV) interactions \cite{Tiwari:2011km} or the conventional VDW interaction \cite{Behera:2022nfn}. Therefore, we use the IHRG parameters throughout this work.

The expression for the proton number fluctuation given in Eq. (\ref{suscep}) is modified from ideal HRG owing to the van der Waals interaction. The baryon chemical potential $\mu_{B}$ is reduced to modified chemical potential $\mu_{B}^{*}$ as defined in Eq. \ref{eq17}. Similarly, the pressure and number density are also modified as in Eq. \ref{eq12}, and Eq. \ref{eq19}. This leads to a large deviation in the higher-order cumulants in MVDWHRG from ideal HRG, as will be shown in the next section. We have used Eq. \ref{paramet1} and Eq. \ref{paramet2} for the $\sqrt{s_{NN}}$ dependencies of the desired ratios in Eq. \ref{ratios}.
\section{Results and Discussion}
\label{res}
It is expected that the medium formed in relativistic heavy ion collisions undergo particlization at the chemical freeze-out transforming into a hadron resonance gas before further evolution towards the detectors. In mapping the QCD phase diagram, conserved charge fluctuations have been proposed and explored as a method to pinpoint the existence and location of the QCD critical point. Net-proton fluctuations, in particular, have been seen as a proxy for the baryon number fluctuations as they constitute the dominant part of the baryons that can be detected. Multiple studies have tried to relate the proton number fluctuations to the baryon number fluctuations within the ambit of various models with methods available in the literature to relate the two~\cite{Fukushima:2014lfa, Vovchenko:2021kxx, Kitazawa:2011wh, Kitazawa:2012at}.

\begin{figure*}[ht!]
\begin{center}
\includegraphics[scale= 0.29]{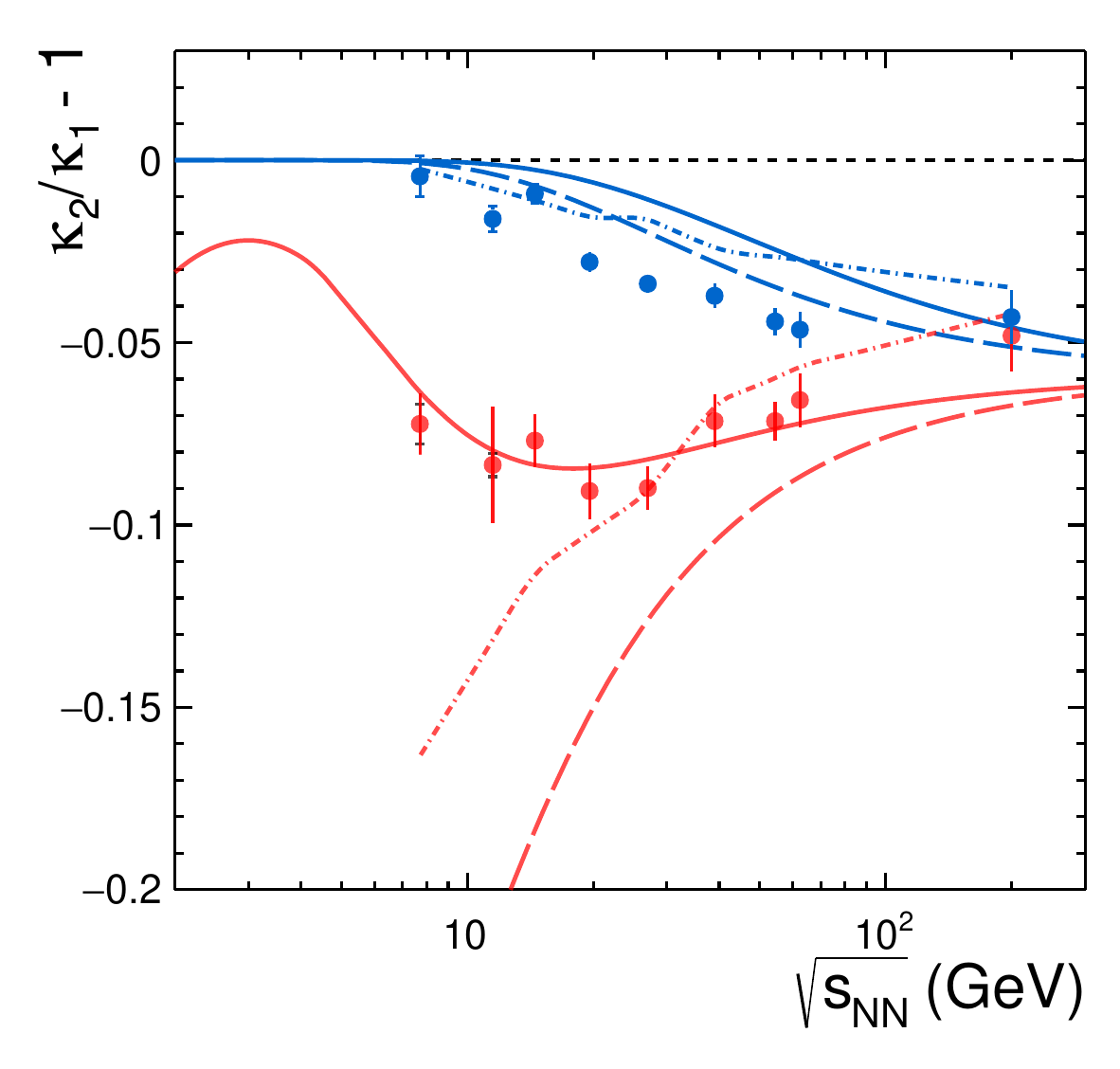}
\includegraphics[scale = 0.29]{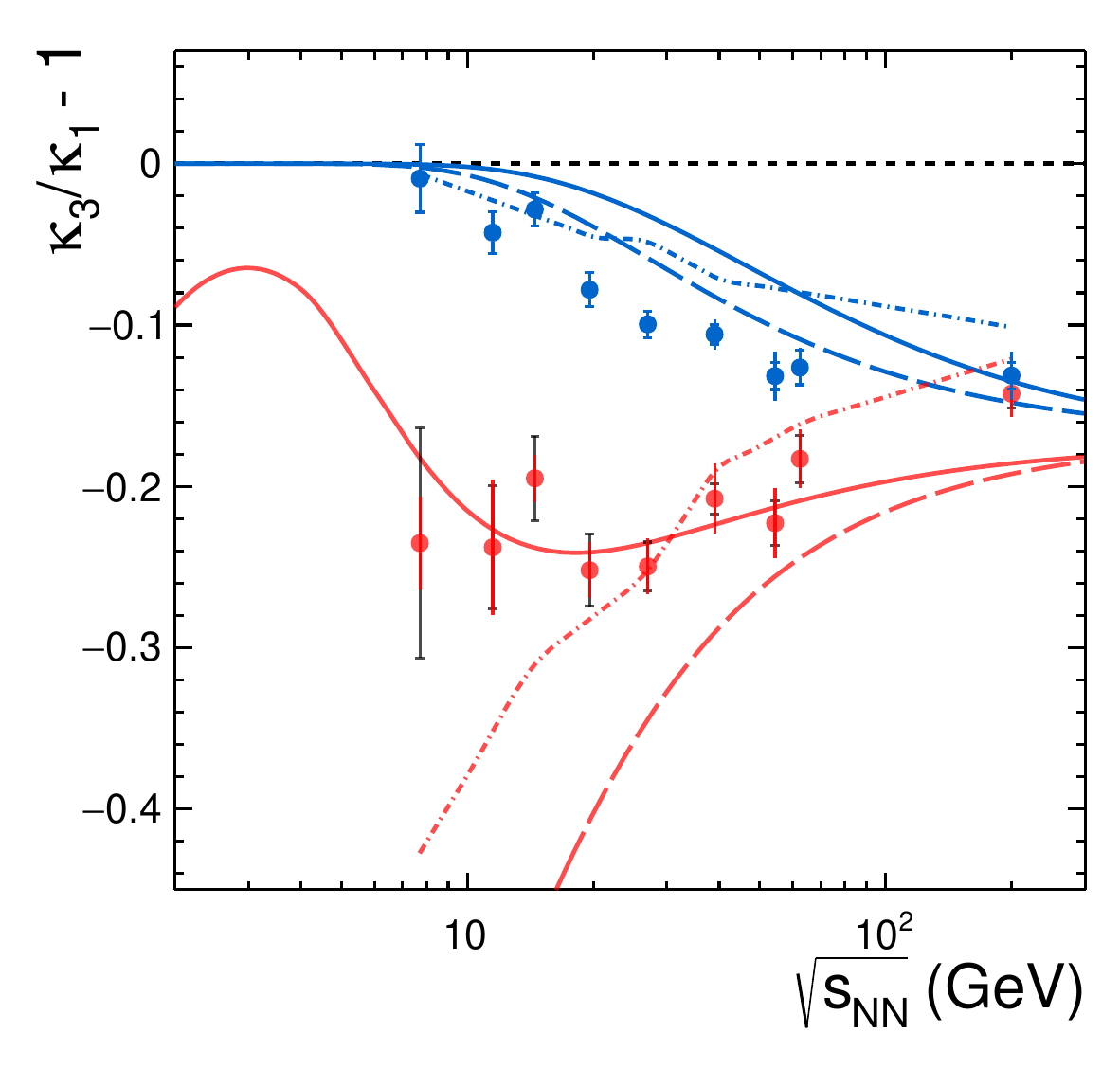}
\includegraphics[scale = 0.29]{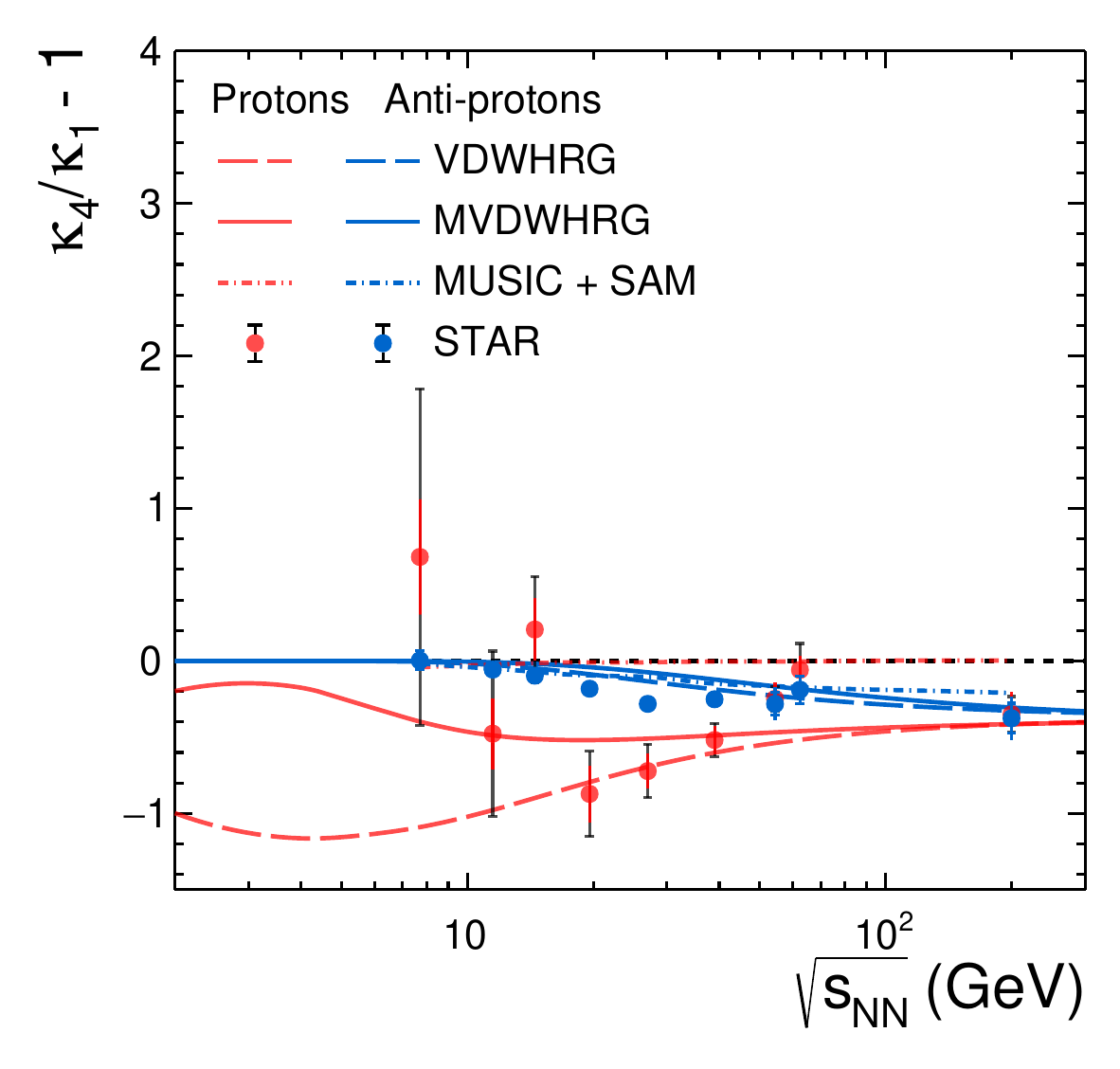}
\caption{(Color Online) Ratios of proton number cumulants as functions of the center-of-mass energy. The red lines and marker correspond to the proton and the blue lines and marker correspond to the anti-proton. The results obtained in MVDWHRG are shown by the solid line, and that of VDWHRG are by the dashed line. The dotted lines are for the results from hydrodynamical calculations \cite{Vovchenko:2021kxx}, whereas, the solid markers are for the STAR data \cite{STAR:2021iop}.}
\label{fig4}
\end{center}
\end{figure*}
\begin{figure*}[ht!]
\begin{center}
\includegraphics[scale= 0.29]{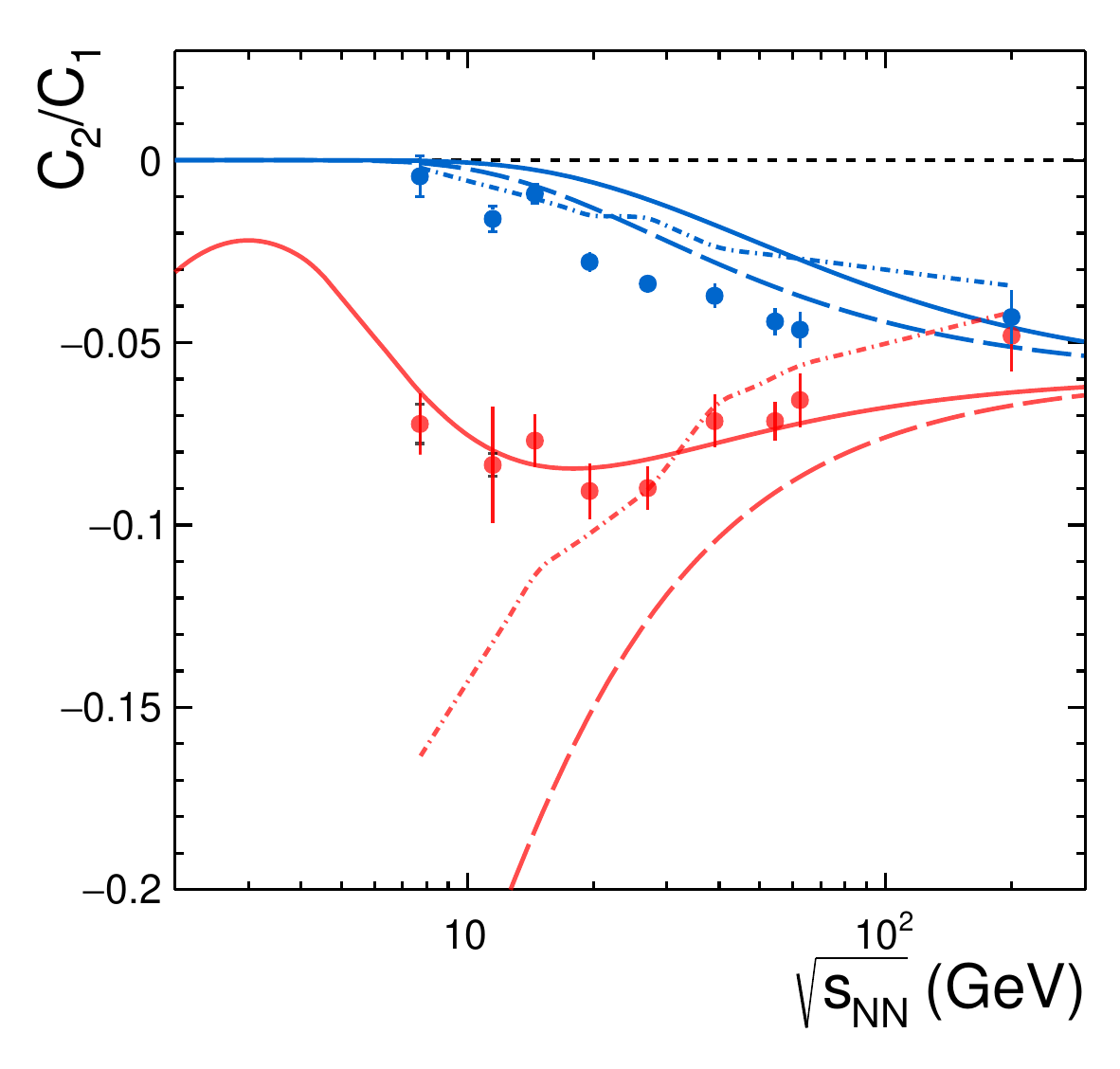}
\includegraphics[scale = 0.29]{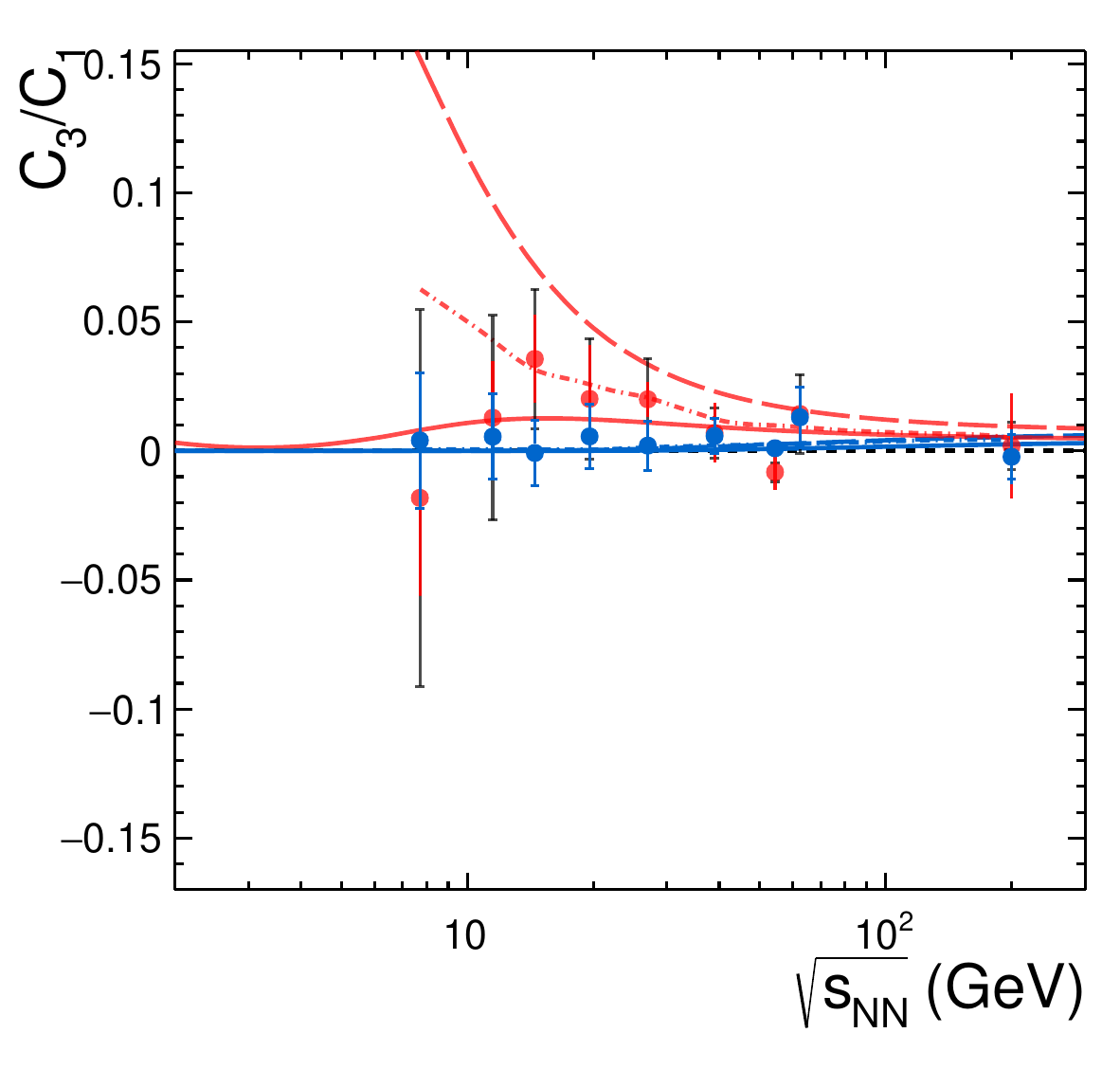}
\includegraphics[scale = 0.29]{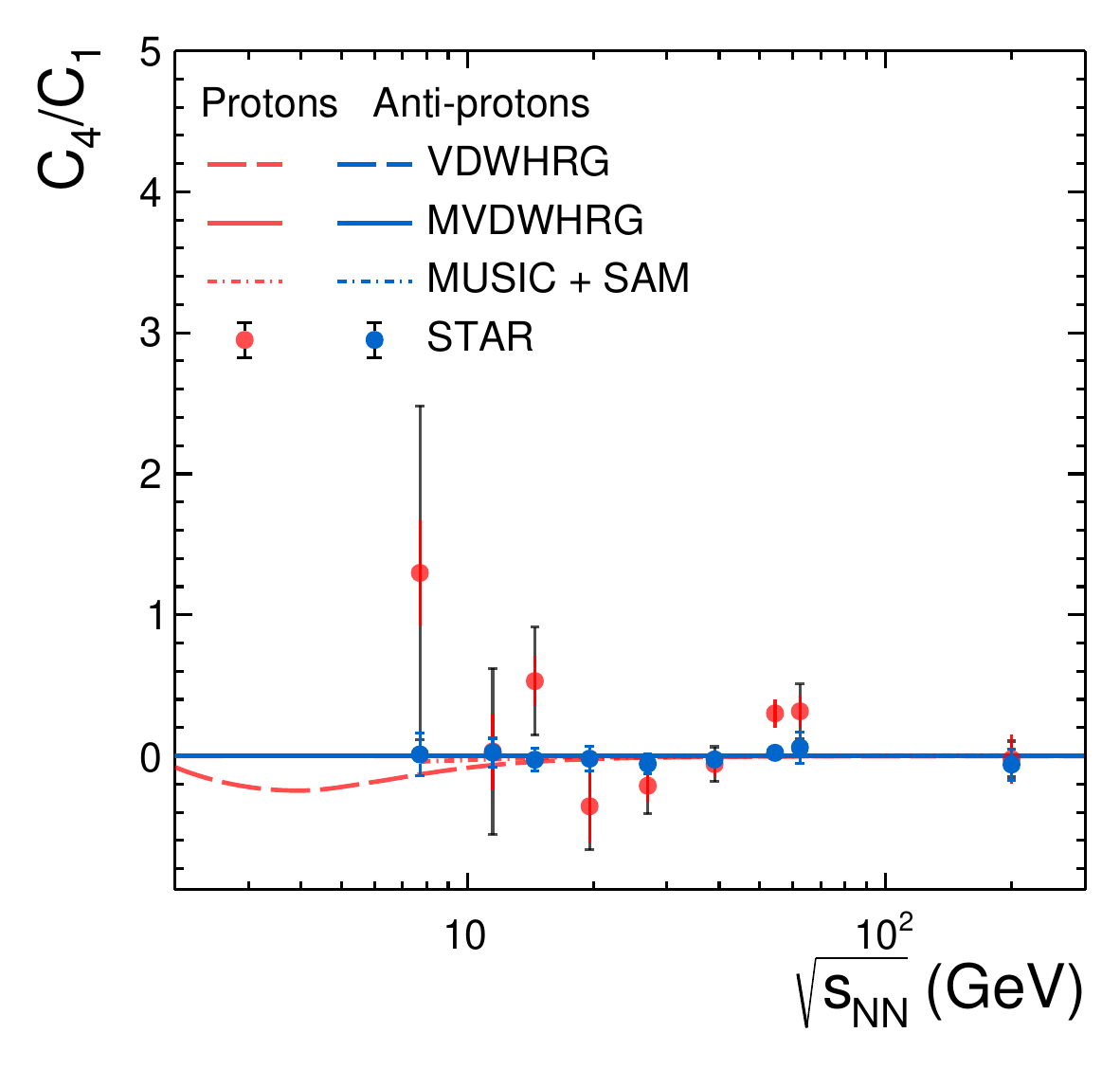}
\caption{(Color Online) Correlation functions as functions of center-of-mass energy. The red lines and marker correspond to the proton and the blue lines and marker correspond to the anti-proton. The results obtained in MVDWHRG are shown by the solid line, and that of VDWHRG are by the dashed line. The dotted lines are for the results from hydrodynamical calculations \cite{Vovchenko:2021kxx}, whereas, the solid markers are for the STAR data \cite{STAR:2021iop}.}
\label{fig5}
\end{center}
\end{figure*}

In principle, the interactions among the hadrons quantify the various dynamics of the hadronic medium. Thus, by considering the attraction and repulsion between the hadrons, one can understand the dynamical behavior of the matter formed in high-energy collisions. In this regard, we have incorporated the baryochemical and temperature dependence of the van der Waals interaction parameters in section~\ref{formulation}, thus appropriately modifying the equation of state. Now, with the modified $a$ and $b$ parameters, we proceed to estimate the net-proton fluctuations as a function of the center-of-mass energy.

In the Boltzmann limit, \textit{i.e.} when the number densities of both $p$ and $\bar p$ are both dilute, one may obtain the Skellam limit for the ratios defined in Eq. \ref{ratios}, 
\begin{equation}
\label{eq.3.2}
    S\sigma = {\rm tanh}~(\mu_B/T), \hspace{0.5cm} \frac{S\sigma^3}{M} = 1, \hspace{0.5cm}   \kappa\sigma^2 = 1.
\end{equation}
Any deviation from this limit may indicate the onset of new physics. The presence of QCD critical point has been predicted to strongly affect these quantities~\cite {Stephanov:2008qz, Stephanov:2011pb, Mroczek:2020rpm}. Fig.\ref{fig3} shows the variation of these parameters as functions of the center-of-mass energy. The solid black markers show the net-proton fluctuation values obtained from the RHIC BES-I experiment \cite{STAR:2021iop}. The black dashed line represents the Skellam predictions, and the solid blue line shows ideal HRG results. The VDWHRG values in solid green are obtained by using the constant VDW parameters, $a$ = 329 MeV fm$^3$ and $b$ = 3.42 fm$^3$~\cite{Vovchenko:2015pya}. All these results are compared with the modified VDWHRG values (solid red), which are obtained by using the functional form of the $a$ and $b$ parameters obtained in section~\ref{formulation}. It was shown in ref.~\cite{Fukushima:2014lfa} that the HRG model deviates from the Skellam condition at high baryon densities due to the inclusion of quantum correlations. But this alone is insufficient to explain the BES-I data for the proton number cumulants, suggesting that these quantum correlations are weak~\cite{Kitazawa:2011wh, Kitazawa:2012at}. The authors in ref.~\cite{Vovchenko:2021kxx} used a hybrid model where further interactions were considered, like the excluded volume effect and global baryon conservation, which were seen to have a significant effect on these ratios. It was suggested that a simple excluded volume component explains the higher center-of-mass energy region well while the deviation below $\sqrt{s_{NN}} \leq 20$ GeV signifies the importance of attractive interactions. The VDWHRG model provides the opportunity to test this but fails in explaining the data, while the MVDWHRG model described in this paper shows a good agreement with the obtained net-proton number fluctuations from BES-I. It can be observed that the deviation from the VDWHRG and MVDWHRG is lesser near the high center-of-mass energies (low $\mu_{B}$). In contrast, the deviation increases with decreasing the center-of-mass energies (high $\mu_{B}$). The interesting thing to note here is that at low center-of-mass energies (high $\mu_{B}$), the MVDWHRG matches with the ideal HRG. This is because at low center-of-mass energy, which corresponds to high values of baryon density, $\mu_{B}$, both of the VDW parameters weaken, approaching zero. Hence the system behaves more like an ideal one.

Further, one may explore the medium by studying the cumulants of protons and anti-protons separately. Correlation functions, $C_n$, may be constructed as a linear combination of ordinary cumulants~\cite{Vovchenko:2021kxx} as given below,
\begin{equation}
    \label{eq.3.3}
    C_1 = \kappa_1
\end{equation}
\begin{equation}
    \label{eq.3.4}
    C_2 = -\kappa_1+\kappa_2
\end{equation}
\begin{equation}
    \label{eq.3.4}
    C_3 = 2\kappa_1 -3\kappa_2 + \kappa_3
\end{equation}
\begin{equation}
    \label{eq.3.5}
    C_4 = -6\kappa_1 + 11\kappa_2 - 6\kappa_3 + \kappa_1.
\end{equation}
These values approach zero for $n>1$ in the Poissonian limit, indicating uncorrelated particle production. Both $\kappa_n$ and $C_n$ were explored at STAR~\cite{STAR:2021iop} as functions of the center-of-mass energy. The comparison of our calculations of these quantities compared with the experimental results is shown in Fig.~\ref{fig4} and~\ref{fig5}. The red and cyan colors represent the results for proton and anti-protons, respectively. The deviations observed from zero indicate the existence of physics beyond the standard uncorrelated hadron gas scenario. The normalized quantities, $\kappa_n/\kappa_1 - 1$ and $C_n/C_1$, are well explained by the MVDWHRG model as compared to the standard VDWHRG model as can be seen from the solid and dashed curves. The results from viscous hydrodynamic simulations that include the combined effect of the excluded volume and baryon conservation were shown to have significant contributions at $\sqrt{s_{NN}}\geq 20$ GeV region as observed in ref.~\cite{Vovchenko:2021kxx}. The change of interaction from repulsive to attractive was suggested as a solution to the deviation observed at lower energies. The inclusion of attractive interactions, which competes with the repulsive interaction through the functional form of $a$ and $b$ is observed to explain the effects being observed in data. These results suggest that using constant $a$ and $b$ obtained by fitting lQCD results at $\mu_B/T = 0$ is not sufficient to describe the system at a non-zero $\mu_B$ adequately. It may also be noted that only a qualitative description of the anti-proton trend is obtained. But, even then, our results are better in agreement with the data as compared with the results obtained in hydrodynamical models~\cite{Vovchenko:2021kxx} or the URQMD model~\cite{Xu:2016qjd,He:2017zpg} which was used for comparison with data in the STAR experiment~\cite{STAR:2021iop}.

\begin{figure}
\includegraphics[scale= 0.42]{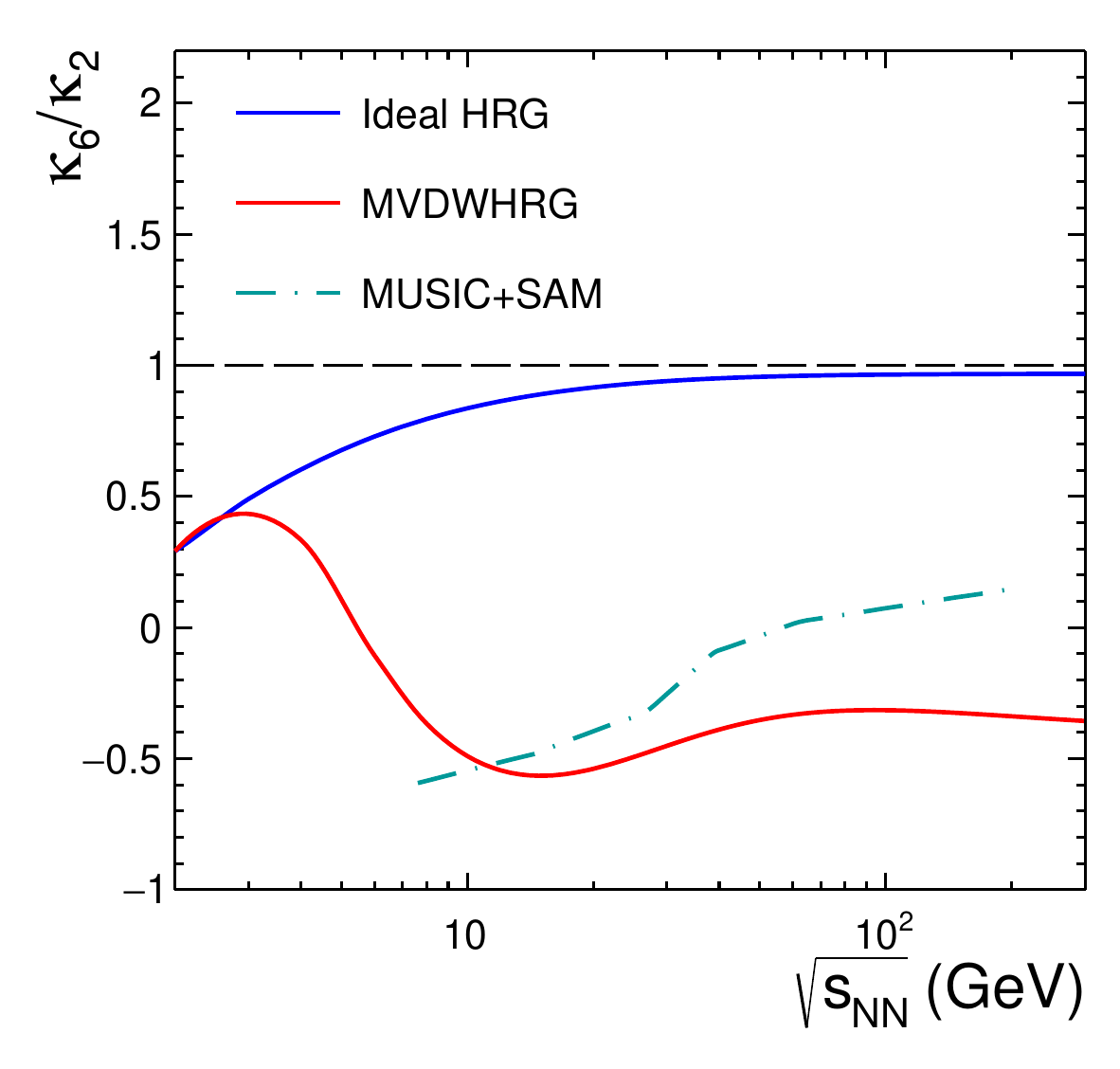}
\caption{(Color Online) Hyperkurtosis as a function of center-of-mass energy. The ideal HRG is shown by a blue solid curve and the results obtained from MVDWHRG is represented by a red solid curve. The dotted-dashed line is for the results obtained using hydrodynamic calculation \cite{Vovchenko:2021kxx}.}
\label{fig6}
\end{figure}

Finally, In Fig. \ref{fig6}, we present our predictions for the net proton hyperkurtosis, $\kappa_{6}/\kappa_{2}$. The result from the ideal HRG shown by the solid blue line is decreasing towards the low center-of-mass energy but remains positive throughout the selected values of $\sqrt{s_{NN}}$. However, the red solid line that represents the results from the MVDWHRG remains negative at the high center-of-mass energy and becomes positive towards $\sqrt{s_{NN}}\leq 5$ and matches with ideal HRG. The prediction for $\kappa_{6}/\kappa_{2}$ from ref.~\cite{Vovchenko:2021kxx} is shown for comparison.

\section{Summary}
\label{sum}

In this paper, we present an estimate of the van der Waals (VDW) attractive ($a$) and repulsive ($b$) parameters obtained from reduced $\chi^{2}$ fits to the lQCD thermodynamic quantities at fixed $\mu_B/T$. The obtained values are then parameterized as exponential functions of $\mu_B/T$. We have further estimated the proton number fluctuations within this modified van der Waals HRG (MVDWHRG) model as a function of center-of-mass energy. It is observed that the MVDWHRG model is able to explain the proton number fluctuation data from the RHIC BES-I experiment. We also give predictions for higher-order cumulants and hyperkurtosis, which is yet to be measured in data.

It is to be noted that the generally expected liquid-gas phase transition, due to VDW interactions, is absent in our model. This is due to the exponential decrease of $a$ and $b$ as a function of $\mu_B/T$, thus approaching the ideal HRG model at high $\mu_B/T$. But as seen from fits to the $\chi$EFT data at $T = 0$ GeV \cite{Fujimoto:2021dvn}, the VDW $a$ and $b$ parameters lie above the lowest values obtained by us on fitting lQCD results. This indicates that the VDW parameters must reach a minimum and then rise towards higher $\mu_B/T$ thus deviating from the exponential fit used in this work and pointing towards possible critical behaviour. But due to the paucity of lQCD results and also of $\chi$EFT data a study on the VDW parameters at these values of $\mu_B/T$ is at present not possible. It would thus be worthwhile and exciting to look for novel approaches to reconcile these first principle models, thereby allowing a complete study of the $T-\mu_B$ plane.

\section*{Acknowledgement}

KP and Ronald Scaria acknowledge the financial aid from UGC and CSIR, Government of India, respectively. The authors gratefully acknowledge the DAE-DST, Govt. of India funding under the mega-science project -- “Indian participation in the ALICE experiment at CERN" bearing Project No. SR/MF/PS-02/2021-IITI (E-37123).

 \end{document}